\documentclass[num-refs]{nbdt-article}

\usepackage{siunitx}
\usepackage[title]{appendix}

\papertype{Original Article}
\paperfield{Journal Section}

\title{Overcoming the Weight Transport Problem via Spike-Timing-Dependent Weight Inference}

\abbrevs{STDWI, Spike-Timing-Dependent Weight Inference; LIF, Leaky Integrate and Fire; RDD, Regression Discontinuity Design.}

\author[*]{Nasir Ahmad}
\author[*]{Luca Ambrogioni}
\author[*]{Marcel van Gerven}

\affil[*]{Department of Artificial Intelligence,
  Donders Institute for Brain, Cognition and Behaviour,
  Radboud University, Nijmegen, the Netherlands}

\corraddress{Nasir Ahmad}
\corremail{nasir.ahmad@donders.ru.nl}

\runningauthor{Ahmad et al.}

\begin{document}

\maketitle

\begin{abstract}
We propose a solution to the weight transport problem, which questions the biological plausibility of the backpropagation algorithm. We derive our method based upon a theoretical analysis of the (approximate) dynamics of leaky integrate-and-fire neurons.
Our results demonstrate that the use of spike timing alone outcompetes existing biologically plausible methods for synaptic weight inference in spiking neural network models.
Furthermore, our proposed method is more flexible, being applicable to any spiking neuron model, is conservative in how many parameters are required for implementation and can be deployed in an online-fashion with minimal computational overhead.
These features, together with its biological plausibility, make it an attractive mechanism for weight inference at single synapses.

\end{abstract}

\section{Introduction}
Backpropagation of error is a successful approach for training rate-based neural network models~\cite{LeCun2015-ck,Schmidhuber2014-ex}.
However, since its inception it has been criticised for its lack of biological plausibility~\cite{Grossberg1987-ok, Crick1989-xf}.
In particular, in order to update individual synaptic connections weights within a network, information is required about distant error signals and the weights of other synaptic connections of the network -- information which is not available locally to the synapse. 
However, backpropagation's flexibility, unrelenting success in application-based research, and most significantly its capacity for modelling and reproducing neural response statistics has contributed to a recent re-examination of its potential role and plausibility in neural systems~\cite{Richards2019-cc,Whittington2019-mw,Lillicrap2019-mi,Yamins2016-wg,Guclu2015-ms}.

A number of attempts have been made to explain mechanisms by which backpropagation's implausibilities can be addressed.
These can be divided into methods which propose alternative implementations of backpropagation, namely energy-based and dynamical systems methods which converge to backpropagation of error~\cite{Whittington2017-wg, Guerguiev2017-zu, Sacramento2018-wb}, for an overview see~\cite{Whittington2019-mw}, and methods which show that components which are considered implausible can be approximated using alternative and plausible computations~\cite{Kolen1994-zy,Lillicrap2016-nm,Akrout2019-az,Guerguiev2019-iu,Kunin2020-zq}.
We focus on the latter approaches in this study.

\begin{figure}[!ht]
    \centering
    \includegraphics[width=0.8\textwidth]{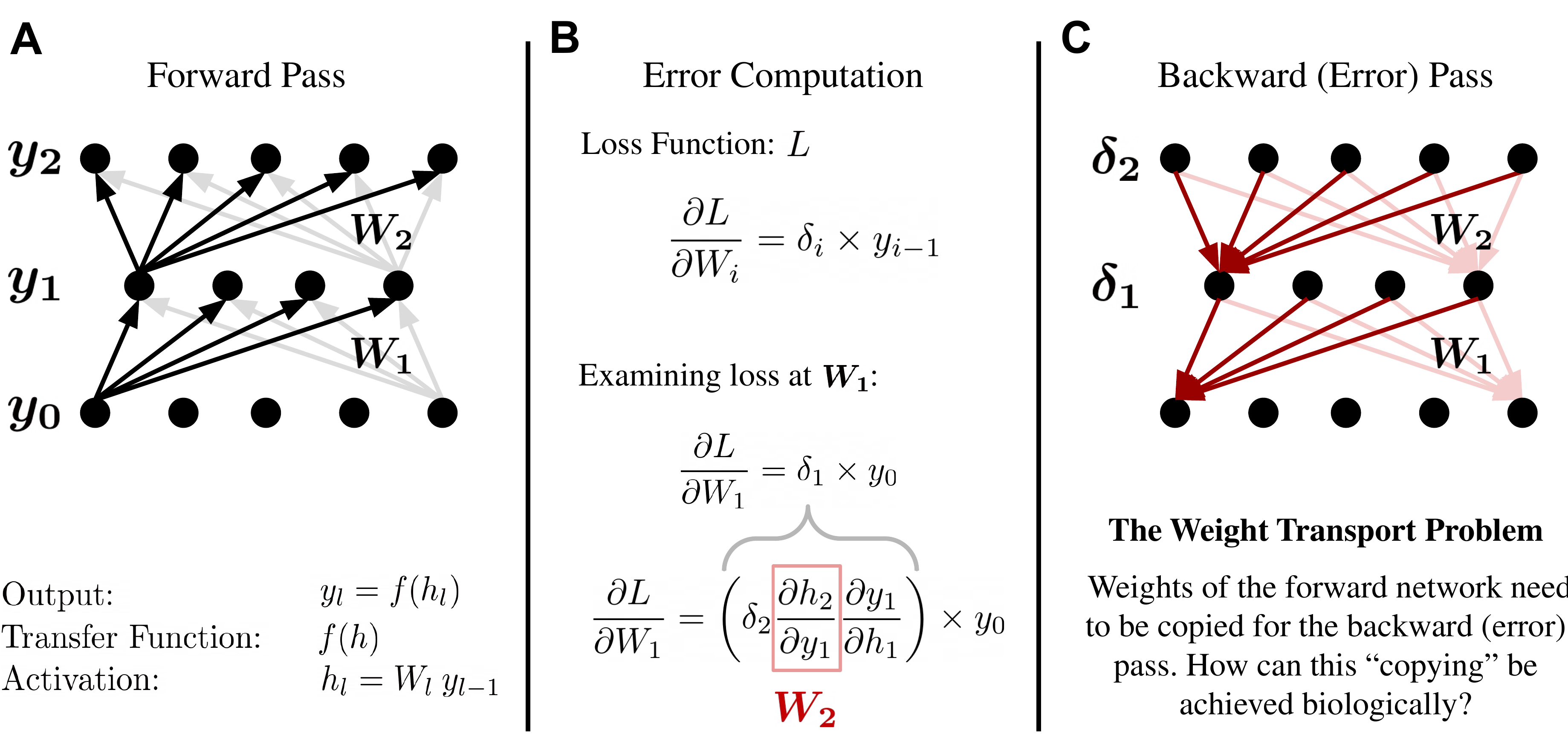}
    \caption{The weight transport problem in backpropagation of error. A. The computations involved in the forward-pass of an example feedforward neural network model. B.  The backpropagation of error method. Specifically, the derivative of the loss function can be computed with respect to each weight matrix in our example network. Observe that the derivative of the loss function with respect to a weight matrix ($W_1$) deep in the network depends upon the weight matrices in the higher layers ($W_2$). C. Backpropagation of error requires a copy of the weights of the forward network.}
    \label{fig:wtp}
\end{figure}

One particularly difficult-to-reconcile component of backpropagation is the need to propagate error signals backwards through a network (see Fig.~\ref{fig:wtp}).
This requires that the backward propagating error signals between layers of neurons is weighted according to the forward synaptic connection weights, leading to a situation in which feedback weight matrices are copies of the feedforward matrices.
This duplication of weights has been identified as particularly troubling in terms of a plausible biological implementation and is known as the {\em weight transport problem}~\cite{Grossberg1987-ok}.

Early attempts to address the weight transport problem included proposals that the feedback weights can converge to the values of the feedforward weights by applying the same weight changes to both matrices during training (see~\cite{Kolen1994-zy}).
This explanation was criticised for simply shifting the problem from transporting weights to transporting weight changes in a network.
More recently, feedback-alignment was proposed as a method to completely sidestep the need for weight symmetry~\cite{Lillicrap2016-nm}.
It was empirically shown that by having fixed random feedback weight matrices between the layers of a network, the feedforward weight matrices are modified by backpropagation such they they come into alignment with the feedback matrices.
This approach can also be implemented with a randomly weighted direct feedback error to every layer (direct feedback alignment,~\cite{Nokland2016-ot}), a method which has also been applied in spiking neural networks~\cite{Samadi2017-eh}.
Though such an error distribution process is biologically plausible, the effectiveness of the approach is limited to shallow networks and the accuracy of deep networks appears to suffer severely under such a protocol~\cite{Bartunov2018-ao}.
Beyond static feedback matrices, matrices with arbitrary magnitudes but alignment of the signs of weights (i.e. positive feedforward weights are mirrored with positive feedback weights and vice versa) show greatly improved performance over feedback alignment~\cite{Moskovitz2018-ae, Xiao2018-qn}. However, propagating the sign of feedback weights is itself a transport problem and performance with this method is less than optimal.

Recently, methods have been proposed by which the symmetric feedback weight matrices could be learned by biologically plausible methods (using local only information).
Specifically, methods have emerged which carry out a process of synaptic weight inference~\cite{Akrout2019-az, Guerguiev2019-iu}.
In essence the backwards synaptic connections (which would propagate the error) attempt to infer the feedforward weight between two neurons by observation of their activity alone.
This is a process in which, based upon the activity patterns of a pair of neurons, a feedback synapse can infer (and thereby copy) the strength of the feedforward synapse.
Such a method was successfully applied in a rate-based neural network by Akrout et al. \cite{Akrout2019-az} (hereafter referred to as the Akrout method).
This method makes use of inference phases during which neurons are randomly stimulated and their activation is correlated in order to infer synaptic weights.
Alternative rate-based methods are available though we do not consider them given their non-locality~\cite{Kunin2020-zq}. 
A more recent proposal~\cite{Guerguiev2019-iu} considers a spiking neural network model and makes use of the spiking threshold of neurons to implement a quasi-experimental, causal inference method known as regression discontinuity design (we hereafter refer to this method as RDD, also see~\cite{Lansdell2019-nt}).
This method similarly uses inference phases in between training epochs in order to infer the backward synaptic weight matrices.

These inference methods have proven successful in inferring the feedforward synaptic weights for use in the feedback weight matrices but also suffer from a number of drawbacks.
First, the Akrout method operates on firing rates and requires a demeaning process which is carried out in batches.
This demeaning and batching process is particularly troublesome when applied to spiking networks where the learning must therefore be carried out offline and firing rates measured by aggregating spikes at specific intervals.
In the RDD method, weight inference requires a piece-wise linear fitting process in order to infer the synaptic weights.
This procedure requires the storage of four times more parameters per synapse (than just the synaptic weight), a second state variable per neuron and a high computational complexity per update.
Though these components and the calculation protocols might be possible for a neuron to compute, they incur a significant computational cost.

To overcome these issues, we propose a spike-timing-dependent weight inference (STDWI) mechanism for solving the weight transport problem in spiking neural networks.
Our method is motivated by analysis of the time-to-spike of various neuron models under the influence of an incident spikes.
In order to estimate this in a biologically plausible and computationally efficient manner, we make use of local information for this computation, in particular just the spike times of the pre- and post-synaptic neurons.
We show that under a number of conditions our method outperforms both the Akrout and RDD methods when applied to weight estimation in spiking neural network models.
We also compare our method to an optimal Bayesian update rule for an integrate-and-fire neuron with stochastic input.
Our rule proves effective as an approximation of this update rule.
Furthermore, for networks in which the neurons emit action potentials at random times (i.e. without a correlation structure), our learning rule can analytically be shown to approximate a rate-based learning rule similar to the Akrout method.
Finally, the update rule we propose is computationally cheap and can be applied in an online fashion.

\section{Methods}\label{sec:Methods}

To address the weight transport problem, it has been proposed that network weights can be inferred from activity~\cite{Akrout2019-az, Guerguiev2019-iu}.
We can formulate this problem as follows: Consider two neurons, labelled ``A'' and ``B'', embedded in a larger network structure.
Amongst other connections, there exists a `forward' synaptic connection from neuron A to neuron B.
Therefore, the activity of neuron B is dependent upon some internal dynamics as well as the network activity as a whole, including the activity of neuron A, via incoming synaptic connections.
Let us also now consider a pseudo synaptic connection from B to A, a connection meant to carry error information backward through the network (note that this work and prior work do not describe this synapse as having an impact upon the network activity during inference).
According to the backpropagation of error algorithm, the optimal value of this synaptic connection weight should be equivalent to the weight of the forward synapse from A to B.
How the forward synaptic weight can be copied to the backward synaptic connection is the problem at hand.

Here we address how to infer the forward synaptic weight value at the backwards synapse given knowledge of the spike times (indexed $k$) of the neurons A and B ($t_A^k$ and $t_B^k$ respectively) and by accounting for some average impact from all other synapses.
We derive a computationally simple and biologically plausible method which, by use of appropriate approximations, achieves this aim and could be employed at the synaptic level to learn feedback weights for error propagation.

\subsection{Derivation of the weight inference method}\label{sec:methods:analytic}
In order to derive our proposed weight inference rule, we analyse a simplified deterministic leaky integrate-and-fire (LIF) neuron with instantaneous synaptic inputs from a single source and drift (where drift is a placeholder for the unknown impact of all other incident synapses) and then consider the impact of noise upon this model.

A deterministic LIF neuron with drift $\mu$ has voltage dynamics
\begin{equation}
\tau_{m}\frac{d v(t)}{ d t } = v_{r} - v(t) + \mu \,.
\end{equation}

In the absence of any input spikes, this equation can be solved, for an arbitrary initial condition $v_0$, at time $t_0$, yielding
\begin{equation}
v(t) = (v_{r} + \mu)\left(1 - e^{-(t-t_0)/\tau_{m}}\right) + v_0e^{-(t-t_0)/\tau_{m}}\,.
\end{equation}
With this expression we can now consider two cases, one in which the neuron is not stimulated by any incoming spikes from neuron $j$ and, beginning at voltage $v_0$ at time $t_0$, it spikes with some time delay $\hat{T}$ (purely under the influence of drift).
The other case is one in which the neuron received an additional instantaneous voltage injection of magnitude $w$ at time $t_0$ (i.e. a spike arrives and stimulates the neuron) and it spikes with a different time delay, $T$ (such that the second case involves replacement of $v_0$ with $v_0 + w$).
These cases can be subtracted at threshold in order to give an expression for $w$, the stimulation magnitude, of the form
\begin{equation}\label{eq:leaky_solution}
    w = {e^{T/\tau_{m}}} (v_{r} + \mu - v_0) \left(e^{-T/\tau_{m}} - e^{-\hat{T}/\tau_{m}}\right).
\end{equation}
Equation~(\ref{eq:leaky_solution}) provides an exact solution for determining the amount of instantaneous voltage ($w$) injected to a neuron at some time $t_0$ given that its spike time was modified from an expected time $\hat{T}$ to the time $T$.
This is under the assumption that other than the instantaneous voltage injection and a background drift, there are no other inputs to the neuron during this time.

We wish to make use of this deterministic solution for application to noisy conditions.
In particular, when the background drift is considered as due to input from many other neurons it would inherently be noisy (unlike our derivation above).
However, the current expression includes a number of terms which are highly susceptible to noise.
First, the exponential term, $e^{T/\tau_m}$ is a strictly positive function which is exponential in the time that the neuron took to spike.
If we consider a case in which $T$ is noisy, this term scales our noise exponentially but never changes sign.
Second, the expected time to spike, $\hat{T}$ is difficult to estimate in a noisy neuron.
However, this term is crucial for our ability to accurately identify positive and negative weights and it must, therefore, be approximated.

First we consider the exponential term $e^{T/\tau_m}$. 
Though this term might (in the noiseless case) aid in producing a highly accurate weight estimation, in the face of noise it introduces a significant error.
Furthermore, in the noiseless case (where a single estimate of the weight is exact), its biggest function is to scale the estimated weight based upon the time taken to spike.
This, in essence, reweighs cases in which the neuron dynamics take excess time to reach threshold -- due to beginning far from threshold (low $v_0$), having a small drift, and/or when the incident spike arrives from a synapse with a small/negative weight.
This is therefore a mechanism to ensure that, for cases in which our system setup results in dynamics which take some time to reach threshold, the weight scale is treated sensibly.
However, in the coming derivation we intend to sample over multiple instances of such an estimation in a noisy system such that there is an unreliable signal of `time to spike'.
And given that this term is heavily influenced by noise we wish to ignore it.
Therefore, given its function, our intention to sample, and its susceptibility to noise, we test in this work the removal of this term from our weight estimation and instead propose weight estimation without this scaling.
We empirically find this approach successful.
Thus, our approach to (approximate) weight estimation can be described as
\begin{equation}
    \tilde{w} = C(v_{r} + \mu - v_0) \left(e^{-T/\tau_{m}} - e^{-\hat{T}/\tau_{m}}\right) 
\end{equation}
where $\tilde{w}$ is an approximate estimation of the weight (ignoring a rescaling based upon time to spike) and we have introduced a general constant $C$ to allow linear rescaling of weight estimates.

Next, we wish to approximate $\hat{T}$ in the face of noisy samples.
For this purpose, let us average our estimate of the weight over $K$ observations.
In particular, let us consider a set of samples $T^k$, indexed by $k$, each of which correspond to the time to spike given that the `output' neuron started from some initial voltage $v_0^k$ at the moment of an incident spike.
For each of these samples, there exists an ``expected'' time from incident spike to neuron spike, $\hat{T}^k$, which corresponds to when the neuron would have spiked if not for this incident spike.
Taking an average of the weight estimate over these $K$ samples yields an estimated weight 
\begin{equation}
    \tilde{w}^K = \frac{C}{K} \sum_{k=0}^K \left(v_{r} + \mu - v_0^k\right)\left(e^{-T^k/\tau_{m}} -e^{-\hat{T}^k/\tau_{m}}\right)
\end{equation}
with $K$ indicating the number of observations/samples taken.
If we assume that our $K$ samples are chosen independently of the incident activity (i.e. the incident spikes are random), then the values of the initial voltage, $v_0^k$, and expected times to spike, $\hat{T}^k$, are both independent of the sampling process (and of $w^k$ and $T^k$).
Therefore, these can be independently averaged and, hence, replaced with $\langle v_0 \rangle$ and $\langle\hat{T}\rangle$.
Thus, we arrive at an expression
\begin{equation}
    \tilde{w}^K = \frac{D}{K} \sum_{k=0}^K \left(e^{-T^k/\tau_{m}} -e^{-\langle\hat{T}\rangle/\tau_{m}}\right)\,,
\end{equation}
where $D = C (v_{r} + \mu - \langle v_0 \rangle)$ combines the various constants and scales our estimate of the weights.

If we now finally consider how we ought to update our estimate of $w$ when we receive an additional $(K+1)$-th sample, we arrive at
\begin{equation}\label{eq:leaky_approx}
    \Delta \tilde{w} = \tilde{w}^{K+1} - \tilde{w}^K = \frac{1}{K+1} \left(D\left(e^{-T^{K+1}/\tau_{m}} -e^{-\langle\hat{T}\rangle/\tau_{m}}\right) - w^K \right).
\end{equation}

Inspecting our derived update rule, the first exponential term in Eq.~(\ref{eq:leaky_approx}) is exponential in the time since an incident spike arrived.
Given this, it is equivalent to sampling a trace which continuously measures the (fast exponential) instantaneous firing rate of the neuron from which the incident spike is arriving.
The second exponential term is exponential in the average time since incident spikes `should' arrive if the weight had been zero, $\langle \hat{T} \rangle$, an measure of the incident spike-rate.
This term can be approximated as a sampling of a slow exponential measure of the average rate of the neuron from which incident spikes arrive.
Finally, the constant term $D = C(v_{r} + \mu - \langle v_0 \rangle)$, has a factor of the drift term $\mu$.
In our model assumption, this drift is background input aside from the synapse under inference and affects the baseline time to spike of our output unit.
This drift therefore scales up with the output neuron's average firing rate.
With these observations, we can make appropriate replacements in order to describe a local spike-timing-dependent weight inference rule.

\subsection{Spike-timing-dependent weight inference}\label{sec:STDWI}
We propose a spike-timing-dependent rule for the purpose of weight inference (STDWI) which can be deployed for parallel and online updates with minimal computational complexity.
Our method maintains multiple online estimates of neuron firing rates through eligibility traces~\cite{Izhikevich2007-le,Gerstner2018-yu} and makes use of these for synaptic weight estimation.
In particular, each neuron (indexed $j$) maintains a fast trace $\epsilon_j^{f}(t)$  and a slow trace $\epsilon_j^{s}(t)$.
The dynamics of the fast and slow traces are calculated for each neuron as 
\begin{equation}\label{eq:STDWI_traces}
\tau_{f} \frac{d\epsilon_j^{f}(t)}{d t} = - \epsilon_j^{f}(t) + S_{j}(t) \quad\text{and}\quad
\tau_{s}\frac{d \epsilon_j^{s}(t)}{d t} = - \epsilon_j^{s}(t) + \frac{\tau_f}{\tau_s}S_{j}(t)\,,
\end{equation}
where $\tau_{f}$ and $\tau_{s}$ are the decay constants of the fast and slow traces respectively, and $S_{j}(t)$ is the spike train of the $j$th neuron.
These traces are computed during simulation in a time-stepping method with the analytic (exponential) solution to these traces computed at every timestep.
This spike train is computed from the set of $k$ spike times of the $j$th 
neuron, $t^{k}_{j}$, such that $S_{j}(t) = \sum_{k}\delta(t - t^{k}_{j})$, where $\delta(\cdot)$ is the Dirac delta function.
Note that these two traces have an equal area (across time) when they both start with an initial value of zero due to the normalization of the slow trace spike-train, scaling factor $\tau_f / \tau_s$.
This property ensures that both eligibility traces act to measure the firing rate of the neurons with the same scale. 
Having defined these eligibility traces, we define our weight inference rule as
\begin{equation}\label{eq:STDWI}
\frac{d w_{ji}}{ d t}  =  \alpha S_i(t)\left(\epsilon_i^{s}(t)\left(\epsilon_j^{f}(t) - \epsilon_j^{s}(t)\right) - \eta w_{ji}\right)\,,
\end{equation}
where this rule describes inference of the weight of the forward synapse at the backward synapse (from neuron $i$ to neuron $j$), $w_{ji}$, with $\alpha$ as the learning rate and $\eta$ as the relative level of weight decay (both constant hyper-parameters).
This learning rule and the fast and slow measures of the neuron's firing rates are inspired by the synaptic inference rule derived in Section~\ref{sec:methods:analytic}.
Note that though this rule is given as a differential equation, since updates are gated by neuron $i$'s spike times, it is implemented as updating the synaptic weights such that $w_{ji} \gets w_{ji} + dw_{ji}/dt$ at every timepoint where neuron $i$ spikes.
\begin{figure}[!ht]
    \centering
    \includegraphics[width=1.0\textwidth]{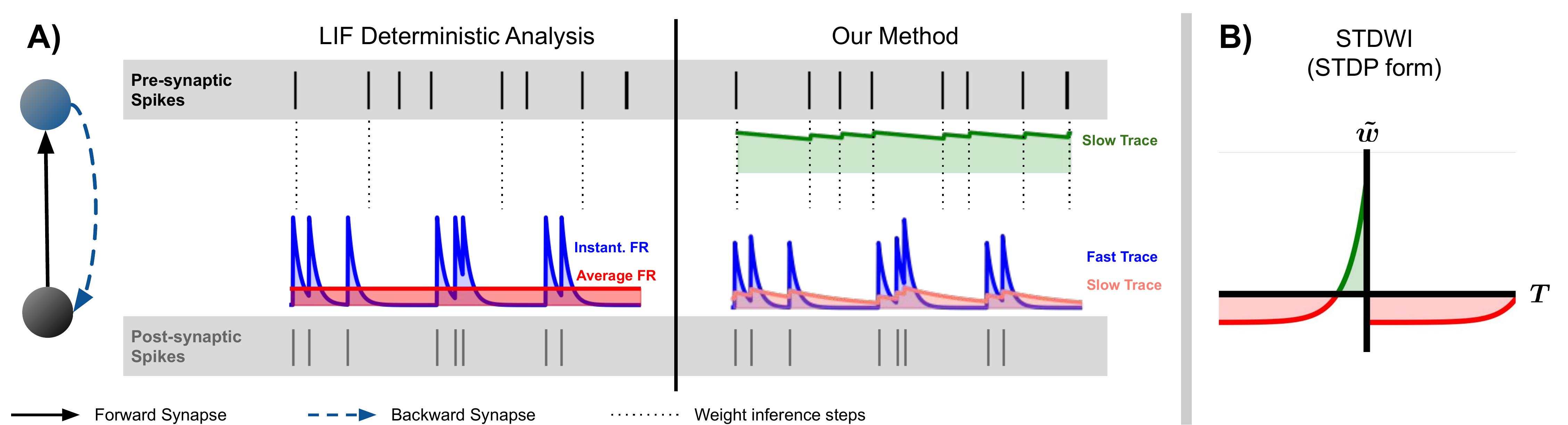}
    \caption{A) Illustration of the difference between our derived method for weight inference by analysis of a deterministic LIF neuron (left) versus our proposed STDWI method (right) which uses a fast trace to measure the instantaneous firing rate (the first exponential term in Eq.~(\ref{eq:leaky_approx})) and a slow trace to measure the average firing rate (second exponential term in Eq.~(\ref{eq:leaky_approx})). B) Assuming regular neuron firing conditions, our method can be interpreted as an STDP rule of the form shown inset, where $T$ is the post minus pre-synaptic neuron spike time. Note pre and post-synaptic are termed relative to the backward synaptic connection.}
    \label{fig:methods:traces}
\end{figure}

The formulation for the weight update given in Eq.~(\ref{eq:leaky_approx}) and our proposed STDWI rule given in Eq.~(\ref{eq:STDWI}) have corresponding terms, see Figure~\ref{fig:methods:traces}A.
Both of these formulations include updates which occur only upon our pre-synaptic neuron spikes.
Note we use the terms post-synaptic/pre-synaptic relative to the backward synaptic connection for application to the weight transport problem.

In our approximation, we replace the first exponential term of Eq.~(\ref{eq:leaky_approx}) (an exponential measure of the time since the post-synaptic neuron's last spike) with a fast timescale measure of the post-synaptic neuron's firing rate (the fast trace) and we use a slow timescale measure of the post-synaptic neuron's firing rate (the slow trace) to approximate the second exponential term (which computes a trace tracking the average post-synaptic neuron's firing rate).
Finally, we include a slow measure of the pre-synaptic neuron's firing rate as a multiplicative factor, which is intended to capture the dependence of the weight estimate upon the pre-synaptic neuron drift.
Figure~\ref{fig:methods:traces}A depicts how updates are calculated upon pre-synaptic neuron spikes for both the deterministic LIF and STDWI update, highlighting both the similarities and key differences between these implementations.

Note that the learning rule being proposed here relates in a curious form to traditional Spike-Timing Dependent Plasticity (STDP) rules.
In particular, the sign of the weight update is determined by the spike-timings and firing-rates of the pre and post-synaptic units.
In general, if we assume some fixed regular firing rate of the post-synaptic neuron, then depending upon the spike-time of the pre-synaptic neuron relative to this regular firing, we encounter positive or negative weight estimation.
This rule therefore appears in a mirrored-form to the commonly cited STDP observations \cite{Bi1998-pi}, see Figure~\ref{fig:methods:traces}B.

\subsection{Spiking neuron model}\label{sec:methods:lif_model}

For simulations in this study, we consider neurons with membrane leakage and conductance-based synaptic kernels whose membrane voltage dynamics can be described by
\begin{equation}
\tau_{m}\frac{d v_i(t)}{d t} = (v_{r} - v_i(t)) + \frac{g_D}{g_L} \bigg(\sum_{j}w_{ij} \kappa_{j}(t) - v_i(t) \bigg)\,,
\end{equation}
where $\tau_{m}$ is the leakage time constant, $v_r$ is the rest voltage, $g_D$ and $g_L$ are the dendritic and somatic leakage conductances, respectively, $w_{ij}$ is the weighting of the forward synaptic connection from the $j$th neuron to the $i$th neuron and $\kappa_{j}$ describes a filtered form of the $j$th neuron's spike train.
The form of the synaptic filtering kernel is taken as a double exponential with a fast rise and slow decay, such that
\begin{equation}
\kappa_{j}(t) = \frac{1}{\tau_{2} - \tau_{1}} \sum_{k} H(t-t^{k}_{j}) \left(e^{-\frac{t-t^{k}_{j}}{\tau_{2}}} - e^{-\frac{t-t^{k}_{j}}{\tau_{1}}} \right)\,,
\end{equation}
where $\tau_{1}$ and $\tau_{2}$ are the timescales of the fast rise and slow decay, taken to be 3ms and 10ms respectively, and $H(\cdot)$ is the Heaviside step function.

When the membrane voltage, $v_{i}(t)$, reaches a threshold, $\theta$, an action potential is recorded and propagated.
The membrane voltage is thereafter reset to a reset voltage $v_{\text{reset}}$. 
For the simulations in this study, we do not implement a refractory period explicitly.
This is not expected to cause much deviation of the analysis in our low firing-rate regime.

\subsection{Comparison against alternative weight inference methods}

In Section~\ref{sec:results:altmethods}, we compare our method (STDWI) to alternative methods (RDD and Akrout methods) proposed for the local inference of weights in a network.
The inference is carried out in networks composed of a group of spiking input neurons connected via a single forward weight matrix to a group of spiking output neuron.
The spiking neuron dynamics are equivalent to those used in the work which introduced RDD as a causal weight inference method~\cite{Guerguiev2019-iu}, see Section~\ref{sec:methods:lif_model}.
The network is stimulated by selectively exciting input neurons, collecting the responses of input and output neurons, and applying the set of techniques to these neural data.

During simulation, some percent of the input neurons are randomly sampled every 100ms and these are excited with background random Poisson distributed spike trains (with a fixed positive synaptic connection weight from stimulation nodes to the input neurons).
Every 100ms the input neurons being stimulated are re-sampled.
During this stimulation, non-selected neurons are left unstimulated with zero input.
The STDWI and RDD methods are applied in a continuous form, paying no attention to the 100ms stimulation periods.
The Akrout method was proposed for rate-based neural networks and makes use of a batch-wise de-meaning of firing rates.
Therefore, the 100ms stimulation periods are considered as individual 'stimuli' for the Akrout method, and the firing rates computed for each of these `stimuli'.
These individual stimuli are then grouped into batches (batch-size chosen by grid search) and used to update the inferred weight according to the Akrout method.

The spiking dynamics with stimulation, as described above, were simulated for 2500s.
During weight inference, these 2500s of dynamics were then looped ten times in order to produce a total 25,000s training time.
This looping was necessary due to memory and storage constraints and can be interpreted as ten epochs of training.

All methods were trained with a learning rate of $5\times 10^{-5}$.
This learning rate was chosen by iteratively reducing the learning rate until stable weight inference was clear for all methods.
Conclusions are (and should be) drawn based upon asymptotic performance and not speed given that this hyperparameter was not tuned on a method-by-method basis.
Free parameters were optimized for by measurement of sign-accuracy and Pearson correlation (best average performance) using a grid search carried out with a single seed of the network simulation.
Selected parameters were then applied for inference to five other network seeds, and results collected.
See Appendix~\ref{app:baseline} for the grid search results. 

\section{Results}

To validate our approach, we compare it against a Bayes-optimal method for a simple neuron model that affords an analytical solution. Furthermore, we compare it to two state-of-the-art synaptic weight inference methods for estimation of the connectivity of simulated spiking neural networks (see models described in Section~\ref{sec:methods:lif_model}). Code to reproduce results is available at \url{https://github.com/nasiryahm/STDWI}.

\subsection{Comparison of STDWI to a Bayesian optimal method}\label{sec:results:bayes}

\begin{figure}[!ht]
    \centering
    \includegraphics[width=0.75\textwidth]{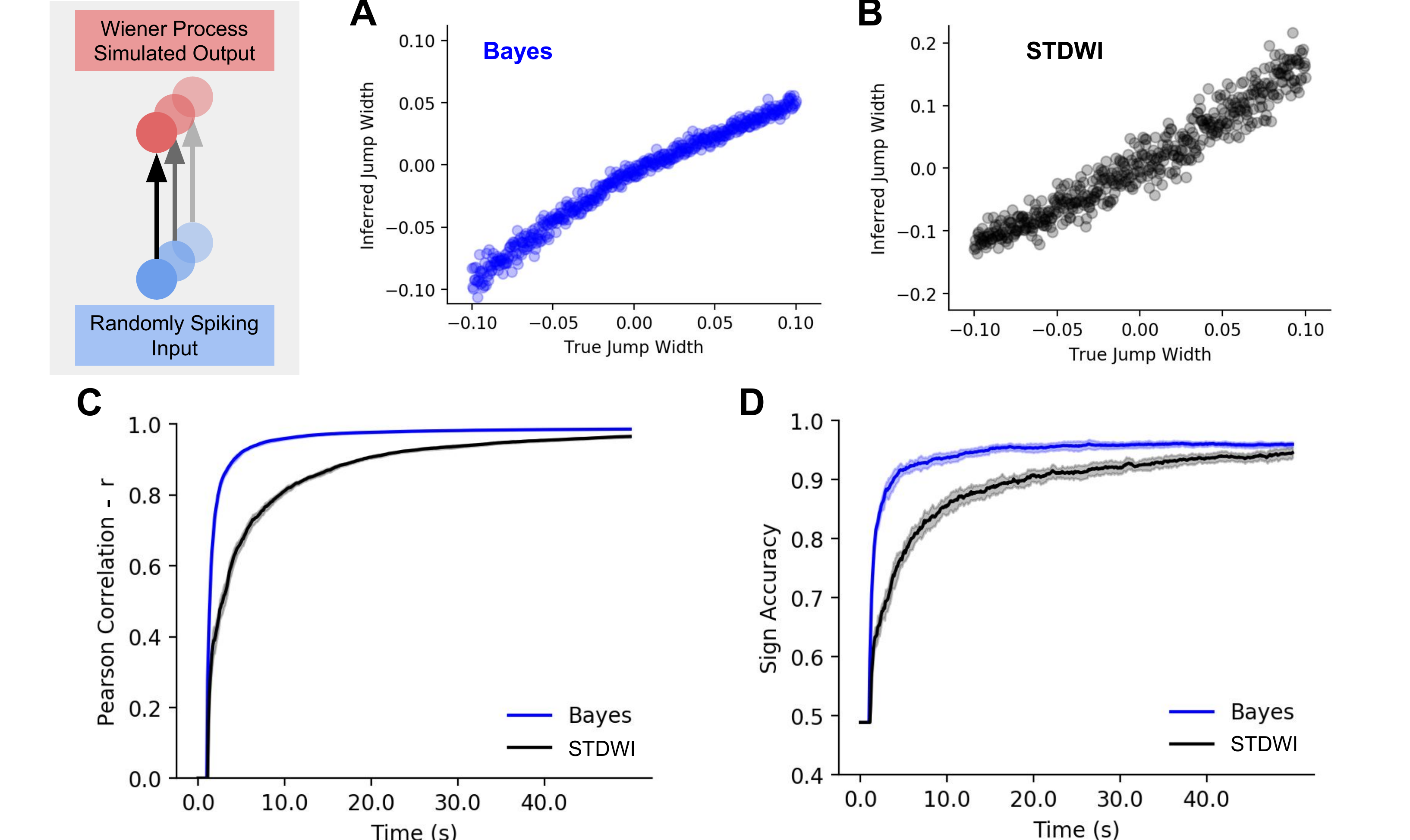}
    \caption{Weight inference accuracy of Bayesian and STDWI approaches applied to a pure Wiener process with jumps.
    Panels A and B show scatter plots of the true and inferred weights for the Bayesian and STDWI approach, respectively, at the end of the training time ($t=50s$).
    Panels C and D show how Pearson correlation and sign alignment between the true and inferred weights evolve through the training process.
    The standard deviation of the measures across 10 random network seeds are shown as envelopes about the curves.}
    \label{results:fig:bayes}
\end{figure}

To verify the validity of our proposed STDWI rule and demonstrate its flexibility, we compare it against a Bayes-optimal method for inferring synaptic inputs to a neuron with internal state modelled by a Wiener process (Figure~\ref{results:fig:bayes}). Unlike a stochastic LIF neuron model, this model has a tractable hitting-time analysis and thereby we can form a Bayesian update rule for estimating the size of a synaptic input given a subsequent output neuron spike time. A detailed derivation of the Bayes-optimal method is provided in Appendix~\ref{app:bayes}.

The Bayesian update rule occurs upon every pre-synaptic neuron spike and is based upon knowledge of when the last post-synaptic spike occurred (rather than knowledge of all past post-synaptic spikes), it would be an improper comparison to test the optimal Bayesian method against our full STDWI rule (which makes use of all previous spikes in its eligibility traces).
Therefore, to ensure a fair comparison we modify our STDWI rule (Eq. \ref{eq:STDWI}) to use only single spikes.
To do this we replaced the slow eligibility traces, $\epsilon_j^{s}(t)$, with a constant (optimally set as the average of the fast traces), and replaced the fast trace, $\epsilon_j^{f}(t)$, with a term which is exponential in the time since the last spike alone (rather than a decaying trace of all past post-synaptic spikes).
This modification is equivalent to Eq.~\ref{eq:leaky_approx} if we treat the second exponential terms as a constant and use an arbitrary learning rate.

We repeatedly simulated stochastic neurons, each with a single forward synaptic input connection but with varying synaptic connection strengths across simulations.
We simulated the systems for 50s and thereafter used the network activity in this time period for synaptic weight inference.
We repeated this analysis for synaptic weight strengths over a wide range to attempt inference of many different synaptic strengths.
Figure~\ref{results:fig:bayes} shows various measures of the similarity between the true and inferred jump widths for this simulation when using either the Bayesian or our derived method for weight inference. Both the scatter plots and learning curves show that the STDWI method closely matches the Bayes-optimal results, supporting the theoretical soundness of our approach.

\subsection{Comparison of STDWI to alternative weight inference methods}\label{sec:results:altmethods}
\begin{figure}[!ht]
    \centering
    \includegraphics[width=1.0\textwidth]{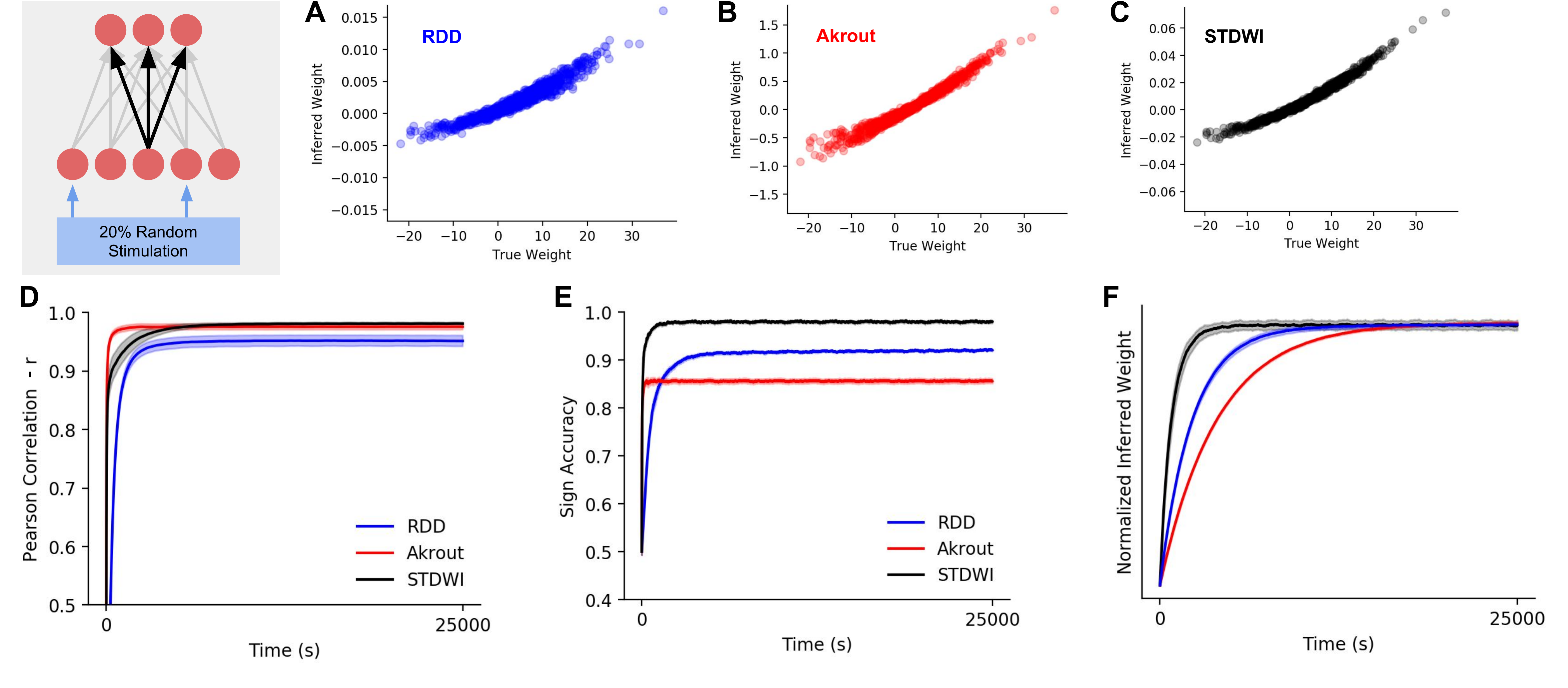}
    \caption{Weight inference accuracy comparison between the RDD, Akrout, and STDWI approaches for a network of LIF neurons with conductance-based synapses.
    Panels A, B and C show scatter plots of the true and inferred weights for each method at the end of training for a single network. 
    Panels D and E show Pearson correlation and sign alignment 
    between the inferred and true weights. Solid lines show the mean of these measures across ten randomly seeded networks and the shaded areas show the standard deviation across these networks.
    Panel F shows the convergence of the inferred weights for each method. The inferred weights for the 75\% largest inferred weight (by magnitude) were collected, individually normalized to their final value and their average plot with standard deviation shown, as before, by the shaded area.}
    \label{results:fig:20perc}
\end{figure}

We also compared our proposed STDWI approach to two existing methods for synaptic weight inference. In particular, we compare against the RDD and Akrout methods. Details of both methods are provided in Appendix~\ref{app:baseline}.

To simulate a neural network model which is amenable to all of these weight inference methods, we use the same neural network models and setup as that described in~\cite{Guerguiev2019-iu}.
This network is composed of LIF neurons with kernel-filtered, conductance-based synaptic inputs.
We simulate two-layer network models with an input layer of 100 LIF neurons fully connected to an output layer of 10 LIF neurons.
The synaptic weight matrix connecting these is drawn from a normal distribution with a small but positive mean.
It is this weight matrix which must be inferred by the range of methods.

The network is stimulated by selectively exciting input neurons.

Some percentage of the input neurons are randomly sampled every 100ms and these are excited with background Poisson distributed input spike trains (with a fixed positive synaptic connection weight from stimulation nodes to the neurons).
Every 100ms the input neurons being stimulated are re-sampled.
During this stimulation process, non-selected neurons are left unstimulated with zero input.

Figure~\ref{results:fig:20perc} shows the result of weight inference with the range of methods discussed above for networks in which 20\% of input neurons are simultaneously stimulated (sparse random stimulation).
Scatter plots of the inferred vs true weights (see Panels~\ref{results:fig:20perc}A-C) show the strength of the STDWI method, which produces a tighter distribution of weights than competitors. Note that the scale of the synaptic weights inferred differs from the true weights for all methods, relating to the approximate nature of the methods.
In practice, a rescaling could be applied to improve the correspondence to the scale of the true weights though none of our measures were sensitive to inferred weight scale and therefore this was disregarded.

Panels~\ref{results:fig:20perc}D and E show the evolution of the Pearson correlation and sign alignment between the true and inferred weights through training for the range of methods.
As can be seen, our proposed STDWI method outperforms both the RDD and Akrout methods, though the difference in the Pearson correlation of all three methods is small. Note that RDD outperforms Akrout in terms of sign accuracy.

Finally, Panel~\ref{results:fig:20perc}F shows the successful convergence of inferred weights for all three methods.
This plot shows the normalized weights (normalised through a division by the final converged weight value) of the top 75\% largest magnitude network weights.
These weights had a relatively unambiguous sign, hence their selection.
This plot is provided to support the argument that the hyperparameter selections made were sufficient for stable inference by these methods.

\subsection{The impact of stimulation protocol on weight inference}

\begin{figure}[ht]
    \centering
    \includegraphics[width=\textwidth]{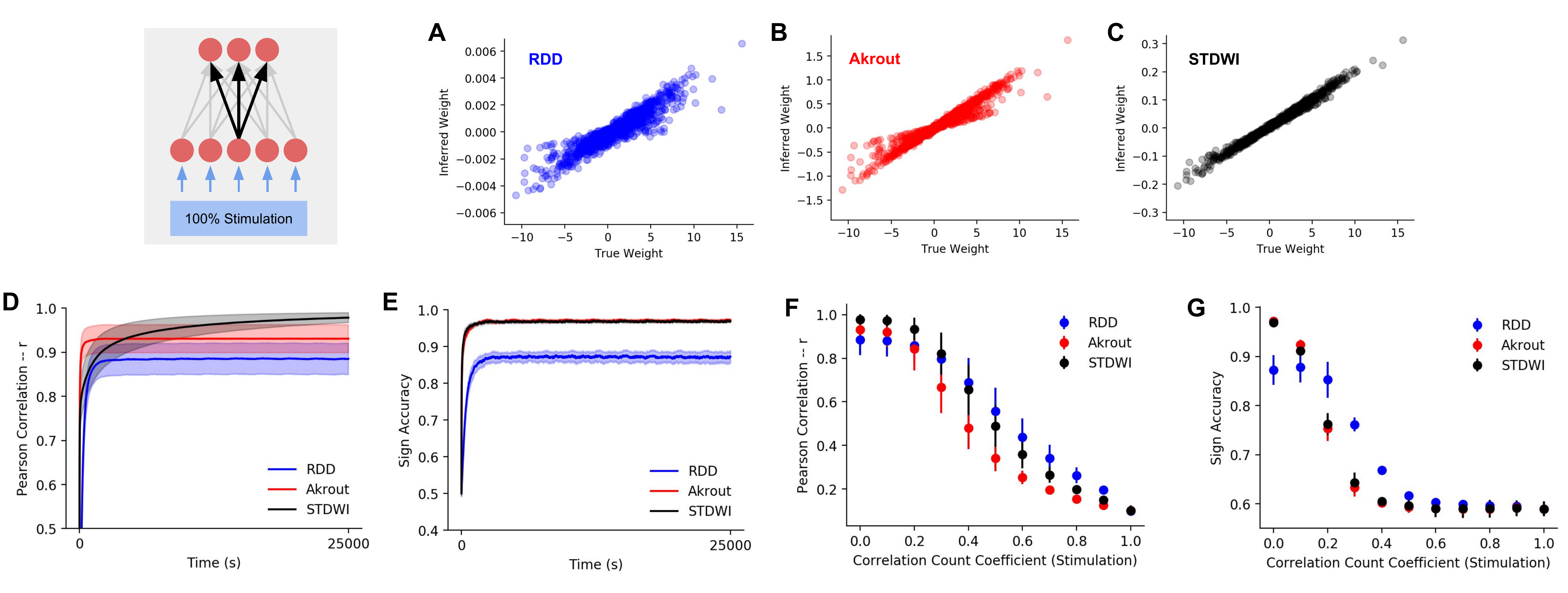}
    \caption{Weight inference accuracy comparison between the RDD, Akrout, and STDWI approaches for a network of LIF neurons with conductance-based synapses when all input neurons are stimulated. 
    Panels A, B and C show scatter plots of the true and inferred weights for each method at the end of training for a single network.
    Panels D and E show Pearson correlation and sign alignment between the inferred and true weights in the uncorrelated spiking case during training. Panels F and G show the final results (post-training) under varying input spike-time correlation. Solid lines (points) show the mean of these measures across five randomly seeded networks and the shaded areas (error bars) show the standard deviation across these networks.}
    \label{results:fig:100perc}
\end{figure}

It is also instructive to investigate how different stimulation protocols affect weight inference.
To this end, and in contrast to the sparse stimulation in the previous section, we assume that all input neurons are stimulated (dense stimulation).
Furthermore, we investigate how input timing correlations affect weight inference.
Since input neurons are stimulated by random Poisson spike trains, we can create correlation between individual Poisson spike trains by a thinning process (via a Single Process Interaction Model, see \citep{Kuhn2003-uc}). Figure~\ref{results:fig:100perc} shows results for this dense stimulation regime.

Scatter plots of the true vs inferred weights (see Panels~\ref{results:fig:100perc}A-C) again show that STDWI produces a tighter distribution of weights than its competitors.
This highlights the smaller impact of stimulation density upon the STDWI inference method compared with the Akrout or RDD methods.
These scatter plots show inferred weights for the dense stimulation case with zero correlation in timing between the various input stimulation spike-trains.

Panels~\ref{results:fig:100perc}D and E show that the STDWI method remains most successful (as measured by the Pearson correlation and sign alignment mean) when compared with RDD and Akrout methods under dense stimulation.
However, the Akrout method benefits significantly from dense stimulation (whereas the RDD method appears to suffer somewhat).
Thus, the RDD method does not systematically outperform the Akrout method as previously reported (cf. Panels~\ref{results:fig:20perc}D and \ref{results:fig:100perc}E).

Panels~\ref{results:fig:100perc}F and G demonstrate how weight inference is affected by input timing correlations.
STDWI remains largely successful, however as input spike timing correlation increases, the RDD method performs favourably.
This may be expected as unlike the STDWI and Akrout methods, the RDD method compares only events which are near-threshold to establish synaptic weights.
This filtering of events by which inference is done may be favourable in the regime of high input spike timing correlation, though the benefit only exists for some parameter range.

\section{Discussion}

Our results demonstrate the efficacy of STDWI for synaptic weight inference across a range of network models and stimulation protocols.
We have shown that our approach successfully approximates Bayes-optimal results in a simple neuron model and outperforms existing methods for weight inference. Our results also highlight the attention that must be paid to the employed stimulation protocols since the efficacy of different synaptic weight inference method has been shown to crucially depend on these.

Existing methods cannot be so indiscriminately applied to arbitrary neuron models. For example, the RDD method requires a neuron model which has a second state variable mimicking the membrane voltage.
This state variable should relax to the same value as the membrane voltage when the neuron is not spiking and otherwise should reflect how ``driven'' the neuron is when it spikes.
However, such a state variable is not necessarily constructable for an arbitrary neuron model.
In contrast, STDWI makes use of spike timing alone and is therefore agnostic to the neuron dynamics being simulated.

In our analyses, we reported both on Pearson correlation and sign accuracy.
STDWI systematically outperformed the alternative approaches on both measures for a range of parameters.
One exception is in the investigation of timing-based correlations (Figure~\ref{results:fig:100perc}F and G), in which RDD outperformed the other methods for inference in the case of the medium correlation regime.
This suggests a particular regime might favour the RDD method, however its failure in other regimes suggests that the current formulation of RDD for analysis about a spike-threshold may not be most effective.

It is also important to realize that the number of variables stored per synaptic connection is greater for RDD compared to either the STDWI or Akrout methods.
RDD requires a fitting process using data-points corresponding to events within which the neuron's membrane voltage was near the spiking threshold.
Aside from the selection of events close to threshold, the RDD method also uses four variables per synaptic connection to characterise a piece-wise linear function (with linear functions fit above and below the spiking threshold).
By comparison, STDWI uses two variables for the fast and slow eligibility traces of each neuron and the Akrout method uses two variables storing the firing rate and mean firing rate of each unit within a batch.

To derive our learning rule, we made use of a deterministic analysis of a LIF neuron and considered the spike times of a single input neuron.
Our deterministic analyses later required approximations in order to remove terms which are highly affected by noise.
Ideally we would instead have carried out a stochastic process analysis for a LIF neuron.
The particular stochastic process to which our leaky neuron model corresponds is known as an Ornstein-Uhlenbeck (OU) process.
Unfortunately a general analysis of the OU process that describes when we ought to expect such a neuron to spike (the hitting time) is non-trivial~\cite{Lipton2018-wt}.
Nonetheless, the efficacy of our assumptions is validated by the quality of our results.
Furthermore, under a rate-based analysis of our proposed STDWI rule, we can show a correspondence to the Akrout rule (see Appendix~\ref{app:ratebased}).

A limitation of our approach is that the inference process considers the spike times of a single input neuron.
Instead, a multivariate approach which would take into account the spike-times of all input neurons to infer synaptic weights could prove even more powerful and accurate for weight inference.
Indeed multivariate analyses, often making use of such multi-neuron spiking information along with cross-correlation measures and statistical significance testing, have been applied previously in approaches which aim to infer neural circuit connectivity from neural data~\cite{Van_Bussel2011-rh,Timme2014-gd,Kobayashi2019-rf,Gerhard2013-px}.
These approaches, however, make use of globally available network information and are not concerned with whether this information is locally available at the synapse.
Instead, we took a simplified but powerful approach which could plausibly be implemented at the single synapse level, providing a candidate solution to the weight transport problem.

Concluding, we have shown that STDWI outperforms existing approaches for solving the weight transport problem. Moreover, it is  more flexible, being capable of application to any spiking network data, while requiring minimal computational overhead. 
The benefits of data efficiency and online computation along with its computational simplicity and accuracy make STDWI a promising biologically plausible mechanism for gradient-based learning in spiking neural networks.

\section*{acknowledgements}
We thank Blake Richards and Jordan Guerguiev for their correspondence and for providing us with the code they used for the RDD method.


\bibliography{main}



\begin{appendices}
\section{Bayesian weight estimation for a stochastic neuron model}
\label{app:bayes}

As a method of verification of our proposed STDWI rule and an exhibition of its flexibility, we compare it against an optimal Bayesian method for inferring a single synaptic input to a neuron with internal state modelled by Brownian motion with drift and diffusion (a Wiener process).
Unlike a stochastic leaky integrate and fire neuron model, this model has a tractable hitting-time analysis and thereby we can form an optimal Bayesian update rule for estimating the size of a synaptic input given a subsequent output neuron spike time.
This synaptic weight inference analysis for this simple neuron model and its similarity to our STDWI rule is described in the following section.

\subsection{Bayesian estimation of synaptic weights}
We wish to estimate the weight of synaptic connection given local-only information. In particular, this involves estimating the weight of a synaptic connection given the spike times of an input and output neuron (input and output relative to the forward synaptic connection) as well as the output neuron's membrane voltages.

Constraining this further, let us estimate a synaptic connection weight, $w$, between two neurons given a single input spike time, $t_{\text{in}}$, and the first output spike time which follows this input spike, $t_{\text{out}}$ where $t_{\text{out}} > t_{\text{in}}$.
If we carry out all analysis relative to the input spike time, $t_{\text{in}}$, we can define the key dependent factors.
First, the output neuron's time to spike (the hitting time), following the input neuron spike, is a key measure which we define as $T = t_{\text{out}} - t_{\text{in}}$.
The initial state of the output neuron is also a determining factor in this analysis as it defines the distance to threshold $\Delta$, which we elaborate on below.
Given this setup and by Bayes' rule, we aim to compute
\begin{equation} \label{eq:bayesP}
p(w \mid T, \Delta) \propto p(T \mid w, \Delta) p(w)\,.
\end{equation}

The likelihood term $p(T \mid w, \Delta)$ can be computed through analysis of the neural dynamics.
To compute it, we must account for the impact of spikes from all other input neurons.
In general this is non-trivial.
To simplify this analysis, we consider the case of a non-leaky integrate-and-fire neuron driven by random input.

\subsection{Stochastic neuron model}
We consider a spiking neural network of neurons with membrane voltage under the effect of Brownian motion.
As such, changes in the membrane voltage, $v(t)$, can be described by 
\begin{equation}
    \frac{dv(t)}{dt} = I(t) \,,
\end{equation}
where $I(t)$ is the total input to the cell at time $t$.
Notably, this change in membrane voltage is agnostic to the current voltage $v(t)$ (meaning there is no leakage effect).
When this membrane voltage meets the threshold, $\theta$, an action potential is emitted and the membrane voltage is directly reset to the reset voltage $v_{\text{reset}}$.

Let us consider the input, $I(t)$, as composed of input from the single synaptic connection and some background stochastic process.
The synaptic connection is modelled as a producing instantaneous voltage injections which occur upon the spike times of the input neurons.
The amplitudes of the instantaneous voltage injections induced by input spikes are equal to the weight of the synaptic connection from input to output neuron, $w$.

Aside from these synaptic inputs, we also consider some background input which is a stochastic process. Assuming that there are a large number of randomly spiking input neurons, we can approximate their impact as a random Gaussian input with some mean and variance.
This describes a stochastic process, known as a Wiener process, with some drift (mean input) and a diffusion constant (variance).
This approximation for a neuron's membrane voltage is valid in the limit of a large number of synaptic connections with small synaptic weight magnitudes.

The above details are all approximations but provide us with a simple description of the neural dynamics such that
\begin{equation}
dv(t) = w \delta(t - t_{\text{in}})dt + \sqrt{D}dX_{i}(t) \,,
\end{equation}
where $\delta(\cdot)$ is the Dirac-delta function and $X(t)$ is a Wiener process with drift $\mu$ and variance scaled by $D$.

\subsection{The hitting time of a non-leaky neuron}

We can now attempt to determine the ``hitting time'' of this system, i. e., the time $T$ at which it makes contact with our neuron membrane voltage threshold.
The hitting-time density for a Wiener process with drift (by which we are approximating our non-leaky neuron) can be calculated as;
\begin{equation}
f(T \mid \Delta) = \frac{\Delta}{\sqrt{2D \pi T^3}} \exp \left(- \frac{(\Delta - \mu T)^2}{2DT} \right) \,,
\end{equation}
where $\Delta = \theta - v_0$ is the membrane voltage distance to threshold ( where $v_0 = v(t_{\text{in}})$), $T = t_{\text{out}} - t_{\text{in}}$ is defined as above, $\mu$ is the drift of our Wiener process, and $D$ is the variance of our Wiener process.
In our neuron model, $\Delta$ corresponds to the difference between some initial membrane voltage $v_0$ and the threshold $\theta$, whereas $\mu$ corresponds to the average input to the output neuron from all input synapses in volts.

The description assumes that the membrane voltage starts at some value $v_0$ and is under constant drift.
However, instead we wish to assume that at the initial time, $t_0 = t_{\text{in}}$, our input neuron fired and added some unknown voltage $w$ to the membrane voltage.
Furthermore, rather than computing a probability distribution over the possible times at which the output neuron might spike, we instead know the next spike time of the output neuron, $t_\textrm{out}$, and wish to use it to infer the weight, $w$.

We can therefore assume that for a given pair of input and output spikes, we have a fixed hitting time, $T$, as described above.
Furthermore, under our synapse description for the non-leaky neuron (where synaptic inputs cause an instantaneous change in output neuron membrane voltage of size proportional to the synaptic weight) our initial membrane voltage, $v_0$, can be represented as the membrane voltage just prior to the input spike, plus the synaptic weight.
That is, we take the limit of $v_0$ from below, i.e., 
$
v_0 = \lim_{t \to t_0^-} v(t) + w.
$
This allows us to augment our first-passage density in terms of $w$ such that
\begin{equation} \label{eq:w_hitting}
f(T \mid w,\Delta) = \frac{(\Delta - w)}{\sqrt{2D \pi T^3}} \exp \bigg(- \frac{(\Delta - w - \mu T)^2}{2DT} \bigg)\,,
\end{equation}
where we now define $\Delta = \theta - \lim_{t \to t_0^-} \: v(t)$.
With this formulation of the hitting-time density, we can compute an estimate of the weight $w$ given a particular set of input and output neuron spike times.
Thereafter we can update our estimate of the synaptic weight of interest through Eq.~(\ref{eq:bayesP}).

To make our inference of $w$ tractable, we first take a Laplace approximation of Eq.~(\ref{eq:w_hitting}).
This produces a Gaussian with mean weight
\begin{equation}\label{eq:bayes_mean}
\hat{w} = \Delta - \frac{\mu T + \sqrt{(\mu T)^2 + 4DT}}{2}\,,
\end{equation}
calculated as the maximum of our likelihood $f(T \mid w,\Delta)$,
and a variance 
\begin{equation}\label{eq:bayes_var}
\hat{\sigma} = 1 / ((\Delta - \hat{w})^{-2} + (DT)^{-1})\,.
\end{equation}

Since we have Gaussian distributions for our likelihood, we can take a Gaussian conjugate prior with mean $\mu_0$ and variance $\sigma^2_0$ and obtain a closed-form solution to our posterior weight when given a single input-output spike pair as
\begin{equation}
{w}_p = \frac{1}{\sigma_0^{-2} + \hat{\sigma}^{-2}} \bigg(\frac{w_0}{\sigma_0^2} + \frac{\hat{w}}{\hat{\sigma}^2} \bigg)\,.
\end{equation}
Similarly, we can compute the posterior variance as
\begin{equation}
{\sigma}_p^2 = \bigg(\sigma_0^{-2} + \sigma^{-2} \bigg)^{-1}\,.
\end{equation}

\subsection{Weight estimation under negligible drift}

Let us assume that the diffusion term, $D$, is sufficiently small compared to the drift $\mu$ (such that $\mu \gg D$).
This allows us to ignore the diffusion term in the numerator of Eq.~(\ref{eq:bayes_mean}).
Having assumed this small diffusion scale, we can then describe the maximum likelihood estimate of the weight as
\begin{equation}
\hat{w} \approx \Delta - \mu T\,.
\end{equation}
Furthermore, recall that $\Delta$ is the distance to threshold when the input neuron spikes, $\Delta = \theta - v(t_\textrm{in})$.
By dividing this distance, $\Delta$, by the drift, $\mu$, we can calculate the expected time of the output spike under drift alone, $\hat{T}$, such that
\begin{equation}
\frac{\Delta}{\mu} = \hat{T} \implies \Delta = \mu \hat{T}\,.
\end{equation}
Given these assumptions, we can approximate Eq.~(\ref{eq:bayes_mean}) as
\begin{equation}\label{eq:approx_bayes}
    \hat{w} \approx \mu\hat{T} - \mu T = \mu(\hat{T} - T).
\end{equation}
This formulation can be understood well if we consider a non-leaky neuron under the effect of drift alone (without any stochastic input) and a single input neuron providing instantaneous voltage injections.
In such a case, with knowledge of the initial membrane voltage and drift of the output neuron, we have a deterministic system which will spike at a specific time, $\hat{T}$.
If we perturb this system with a spike from an input neuron (which causes a jump in the membrane voltage), we can decode the synaptic weight by simply measuring the effect on the timing of the output neuron spike time.
The induced change in the output spike time is linearly proportional to the synaptic weight.

\section{Details on baseline methods}
\label{app:baseline}

The STDWI method is compared to existing methods for synaptic weight inference. We provide more details on these methods below.

\subsection{The Akrout method}
In our simulations of LIF neurons, we compare against the Akrout method~\cite{Akrout2019-az}.
This rate-based method makes use of an inference phase in which neurons are stimulated (with mean zero) and then the levels of activity of input and output neurons are correlated to form a weight estimate.
This approach was shown to be highly successful for weight inference and thereby training of rate-based neural network models.
However, since we simulate spiking neurons, which cannot have a negative firing rate, we instead demean the neuron firing ratesand randomly stimulate the input neurons (post-synaptic from the perspective of the backward synapse).
In particular, we use an update rule of the form
\begin{equation}
\Delta w_{ji} = \eta (r_i - \langle r_i \rangle)(r_j - \langle r_j \rangle) - \eta\lambda w_{ji}\,,
\end{equation}
where $\Delta w_{ji}$ is the update to backward synaptic weight, from a neuron indexed $j$ to a neuron indexed $i$, which is attempting to  estimate the weight of the forward synaptic connection, $w_{ij}$.
$r_i$ and $r_j$ denote the firing rates of the $i$th and $j$th neurons, and $\langle \cdot \rangle$ indicates an average of these over time.
Parameters $\eta$ and $\lambda$ are the learning rate and the weight decay respectively.
The learning rate is fixed at with value $\eta=0.0001$ and the weight decay determined by grid search, see below.
The firing rates $r_i$ and $r_j$ are calculated by computing the firing rates within the non-overlapping 100ms stimulation periods of the network.
These stimulation periods are then grouped into batches (of size again determined by grid search) for calculation of the mean firing rates for this batch ($\langle r_j \rangle$ and $\langle r_i \rangle$ respectively) according to the weight-mirror gradient descent method described in~\cite{Akrout2019-az}.

\subsection{Regression discontinuity design}
We also compare against the regression discontinuity design (RDD) method, which was proposed for application in spiking neural networks~\cite{Guerguiev2019-iu}.
It makes use of all times at which a neuron spiked or almost spiked (i.e. its membrane voltage came within some margin of the spiking threshold but never reached it).
It thereafter separately fits the almost-spiked and spiked events linearly against the membrane voltage.
Notably, for the spiking events, a non-reset version of the membrane voltage is used for the linear fitting.
Following a fitting process, the discontinuity of these linear fits at the spiking threshold is used as a measure of the synaptic weight.
For full details of the RDD implementation, see~\cite{Guerguiev2019-iu}.

\subsection{Grid-based optimization of free parameters}

The methods compared have a number of free parameters that can be optimized for.
In case of STDWI these are the time constants of the fast ($\tau_f$) and slow ($\tau_s)$ traces.
In case of RDD these are the distance to threshold at which samples are initiated and the window duration of a sample.
For the Akrout method, the weight decay scaling and the batch-size are hyperparameters.

These parameters are chosen from a grid-search using a single test network's spike trains.
The parameters producing highest average sign accuracy and Pearson correlation between the inferred and true weights are then chosen for analysis of a further four networks (each with a different random seed for input stimulation and the synaptic weight matrix).

\begin{figure}[ht]
    \centering
    \includegraphics[width=1.0\textwidth]{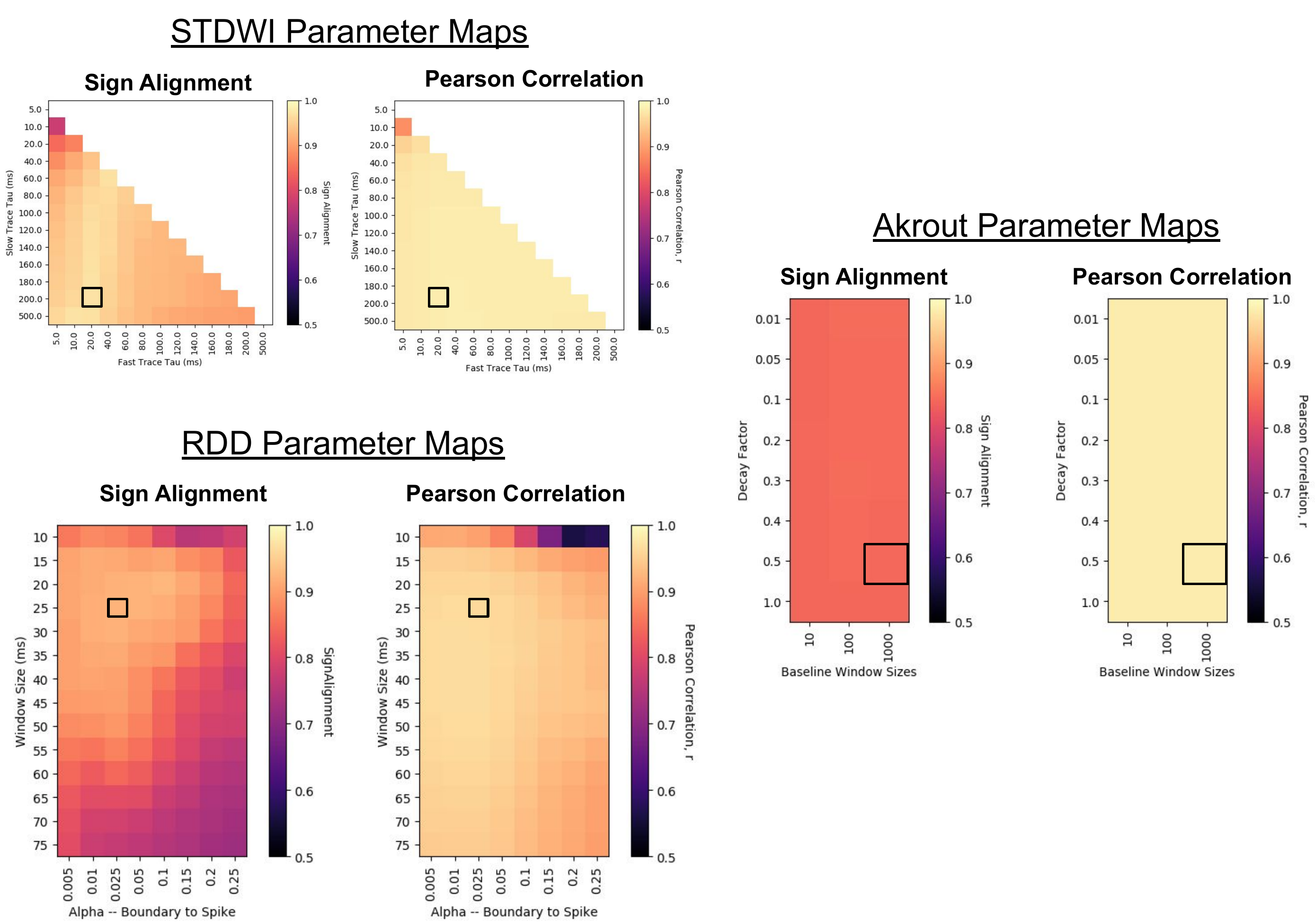}
    \caption{Variation in the performance of the STDWI, RDD, and Akrout methods with changes in the method parameters.
    The best parameter sets are highlighted with a black box. These were the parameter used to analyse all other seeded networks and produce the main results.}
     \label{appendix:parameter_maps}
\end{figure}

Figure~\ref{appendix:parameter_maps} shows the parameters maps for the grid searches carried out to select parameters for Figure~\ref{results:fig:20perc}.
The same grid search parameter sweeps were repeated in order to choose parameters for Figure~\ref{results:fig:100perc}.

\section{Rate-based analysis of the STDWI rule}
\label{app:ratebased}

To appreciate the effect of STDWI rule, we can consider its approximate rate-based form under the assumption of random Poisson process sampled neuron spikes (for a review of the rationale of such methods see~\cite{Morrison2008-mk}).
This produces an update rule based upon the firing rates of the neurons.
Note that below, as in Section \ref{sec:STDWI}, we refer to pre/post-synaptic relative to a `backward' synapse.

In our case, the dependence upon the post-synaptic firing rate has two forms which correspond to a quickly-adapting exponential average, ${\lambda}_\textrm{j}^\textrm{f}$, and a slowly-adapting exponential average, ${\lambda}_\textrm{j}^\text{s}$.
Similarly there is a dependence upon the pre-synaptic firing rate as a slowly-adapting exponential average, ${\lambda}_\textrm{i}^\text{s}$.
Taking the assumption of Poisson random spiking, we can describe our weight update in a rate-based form as
\begin{equation}
\frac{d\hat{w}_{ji}}{dt} = \alpha S_i(t) \left( \lambda_\textrm{i}^\textrm{s} ({\lambda}_\textrm{j}^\textrm{f} - {\lambda}_\textrm{j}^\textrm{s}) - \eta  \hat{w}_{ji}\right).
\end{equation}
We can solve this equation for its fixed point ($\frac{d\hat{w_{ji}}}{dt} = 0$), producing an expression for the fixed-point weight as 
\begin{equation}
\label{eq:STDWI_fixedpoint}
\hat{w}_{ji}^* = \frac{1}{\eta}{\lambda}_\textrm{i}^\textrm{s}\left({\lambda}_\textrm{j}^\textrm{f} - {\lambda}_\textrm{j}^\textrm{s}\right)
\end{equation}
when $S_i(t)$ is non-zero.

For networks with solely positive firing rates, Akrout et al.~\cite{Akrout2019-az} proposed correlating the demeaned firing rates of pre and post-synaptic neurons in order to estimate synaptic weights.
If we here interpret the slow firing rate measure of the input neuron activity as an approximation of its average value, then our method similarly correlates pre-synaptic firing rate with the demeaned post-synaptic neuron firing rate.
Though this rate-based analysis shows similarities to the Akrout method, our spike timing implementation is unique in that it makes use of asymmetric causal kernels and has a demeaning process which slowly tracks the firing rates of neurons (rather than making use of batches).
We attribute our performance gains to these features.
Furthermore, given the spike-timing-dependent feature of the rule, weight updates can be computed in an event-driven fashion and with minimal communication between neurons (weight updates requiring communication only upon spike times).

If we compare Eqs.~(\ref{eq:STDWI_fixedpoint}) and (\ref{eq:approx_bayes}) then we can also appreciate the correspondence of the STDWI rule and the Bayesian estimate.
The STDWI update, instead of making use of an estimate of the a drift, $\mu$, makes use of the pre-synaptic neuron firing rate as a proxy.
This is appropriate given the linear relationship between drift and firing rate for a non-leaky neuron.
Furthermore, rather than directly comparing the expected and true time to spike, $\hat{T}$ and $T$ respectively, the STDWI rule keeps track of a slow and fast estimate of the post-synaptic neuron firing rate, through $\lambda_\textrm{j}^{\text{s}}$ and $\lambda_\textrm{j}^{\text{f}}$ respectively.
The subtraction of these firing rate estimates in Eq.~(\ref{eq:STDWI_fixedpoint}) provides a measure with a similar form to the subtraction of expected and true spike times ($\hat{T} - T$).
Specifically, an earlier than average spike time induces a positive weight estimate and a later than average spike time induces a negative weight estimate.
\end{appendices}

\end{document}